\documentstyle[12pt,aasms4]{article}
\def\vmi{ \ifmmode{(V-I)} \else $($$V$$-$$I$$)$ \fi}
\def\arcsec{\ifmmode{^{\prime\prime}}\else$^{\prime\prime}$\fi}
\def\solar{ \ifmmode {_\odot}\else $_\odot$ \fi }
\def\dg{\ifmmode{^\circ}\else $^\circ$ \fi }
\def\arcsq { \ifmmode { mag/arcsec^2 } \else $\rm mag/arcsec^2$ \fi}
\def\fe{\ifmmode {\rm [Fe/H]}\else {[Fe/H]}\fi}
\def\teff{\ifmmode{\rm T_{eff}}\else${\rm T_{eff}}$\fi}
\def\logg{\ifmmode {\log g}\else ${\rm\log g}$\fi}
\def\mul{\ifmmode {\rm \mu Leo}\else {$\rm \mu$ Leo}\fi}
\def\bww{\ifmmode {\rm BW~IV-167}\else {BW~IV-167}\fi}

\begin{document}

\title{
HIGH-RESOLUTION ABUNDANCE ANALYSIS OF VERY METAL-RICH STARS
IN THE SOLAR NEIGHBORHOOD
\altaffilmark{1}
}

\author{
SANDRA CASTRO
}
\affil{
Instituto Astron\^omico e Geof\'isico, CP 9638
S\~ao Paulo, SP 01065-970, Brazil, and
Dept. of Astronomy,
Columbia University,
538 W. 120th St., New York, NY 10027 
}

\author{
R. MICHAEL RICH
}
\affil{
Department of Astronomy,
Columbia University, 538 W. 120th St.,
New York, NY 10027
}

\author{
M. GRENON
}
\affil{
Observatorie de G\'en\`eve, Chemin des Maillettes 51,
CH-1290 Sauverny, Switzerland
}

\author{
B. BARBUY
}
\affil{
Instituto Astron\^omico e Geof\'isico, CP 9638
S\~ao Paulo, SP 01065-970, Brazil
}
\and 

\author{
J. K. McCARTHY
}
\affil{
Dept. of Astronomy, Caltech, MS 105-24, Pasadena, CA 91125
}
\altaffiltext{1}{
Observations collected at the McDonald Observatory, Texas, USA
}

\begin{abstract}

We report detailed analysis of high-resolution spectra of 
nine high velocity metal-rich dwarfs
in the solar neighborhood, selected from the proper motion
samples of Grenon.    
The stars are super metal-rich, 
and 5 of them have \fe $\geq$ $+$0.4, making them the
most metal-rich stars currently  known.

We find that $\alpha$-elements decrease with increasing metallicity; s-elements are
underabundant by about [s-elements/Fe] $\approx -$0.3; the \ion{Eu}{2} line was measurable for 6 stars, showing [Eu/Fe] $\approx$ 0.0, except for G161-29 where [Eu/Fe] = $+$0.50. 

All calculations followed the same methods used by Castro {\it et al.}
(1996) [AJ, 111, 2439] for the analysis of the well-known metal-rich prototype
\mul\ and the very strong-lined bulge star \bww,
for which \fe =$+$0.46 and $+$0.47 were found respectively.

While exceeding the [Fe/H] of current bulge samples, the chemistry
of these stars has important similarities and differences.
The near-solar abundances of the alpha-capture elements places
these stars on the metal-rich extension of McWilliam \& Rich
(1994) [ApJS, 91, 749], but their s-process abundances are much lower than
those of the bulge giants.  These low s-process values have
been interpreted as the hallmark of an ancient stellar population.
We are unable to convincingly assign these stars to a known
Galactic population and we urge further studies of larger samples.

\end{abstract}

\section{Introduction}

The Galaxy contains a number of little studied old, metal-rich stellar
populations.
In the canonical view of chemical evolution, metals slowly
built up in the interstellar medium over time, and the youngest
disk stars should be more metal-rich than the Sun.  However,
the actual metal-rich populations are instead {\it older} than
10 Gyr, particularly the field population of the Galactic bulge,
and the bulge globular clusters (Ortolani {\it et al.} 1995).  One may add
to this list the metal-rich old open clusters in the disk (Phelps \&
Janes 1996),
and disk dwarfs and giants in the Solar vicinity.    While there
are heuristic explanations for early enrichment in the Galactic
center region, it is more difficult to explain the early formation
of old metal-rich star clusters and local field stars away from
the Galactic center.  Yet high metallicities are found in these
stars, and it is interesting to speculate that they may 
be related to the Galactic center in some way.  On the other
hand, these data may be telling us that high abundances could
be reached in disk regions of relatively lower density, more
distant from the Galactic Center.

About 4\% of local disk stars are super metal-rich
(defined as more metal-rich than the Hyades of 
\fe =$+$0.12), with metallicities
estimated from Geneva photometry in the range $+$0.30 $<$\fe$<$ $+$0.60.
They are either the ultimate stage of the local disk chemical evolution
or genuine members of the Galactic bulge, now scattered out in the disk.

The most metal-rich stars analyzed in the literature are found in the
samples of  Edvardsson {\it et al.} (1993), Luck \& Challener (1995), 
Feltzing (1995) (hereafter F95) for the disk, McWilliam \& Rich (1994)
(hereafter MR94) and Castro {\it et al.} (1996) for the bulge.

Detailed compositions have also been determined for large numbers
of local disk stars, some of which also have high metallicity.
Edvardsson {\it et al.} (1993) analyzed a sample of 189 F and G disk 
stars in the range $-$1.06 $<$\fe$<$ $+$0.26. F95 extended
Edvardsson et {\it al.}'s sample by analyzing 50 disk metal-rich stars
of metallicities in the range $-$0.08 $<$\fe$<$ $+$0.42. Luck \& Challener (1995)
analyzed 55 F/G field giants in the metallicity range of  $-$0.34 $<$\fe$<$ $+$0.39.

MR94 found that 11 Galactic
bulge K giants in the metallicity range of  $-$1.08 $<$\fe$<$ $+$0.44
show enhancement of some $\alpha$-elements such as
Mg and Ti relative to Fe, but not in others such as Ca and Si.
The MR94 giants also had Solar [s-process/Fe] ratios.
Sadler, Terndrup \& Rich (1996) found [Mg/Fe]${\rm \approx}$$+$0.30 for a sample
of 400 K and M giant stars in the Baade Window. Idiart, Freitas Pacheco \& 
Costa (1996) analyzed the integrated spectrum of the Baade Window
and found [Mg/Fe] =$+$0.45 and \fe = $-$0.02, characterizing the bulge population
mix. Castro {\it et al.} (1996) obtained \fe = $+$0.46
for \mul, considered the most metal-rich disk star,
and $+$0.47 for \bww, a strong lined star discovered by
Rich (1988) and the most metal-rich star in the
MR94 sample.

In this work we carry out detailed analysis of a sample of
very metal-rich stars selected on the basis of their kinematics
(perigalactica $\approx$ 3 kpc) and photometry (high metallicities).
The most straightforward expectation is that
[$\alpha$-elements/Fe] $>$ 0 would be
found if these stars formed from rapidly enriched gas (bulge-like
history), 
whereas [$\alpha$-elements/Fe] $\approx$ 0
if they are metal-rich disk stars (Edvardsson {\it et al.} 1993; 
F95).  In reality, the even the MR94 bulge giants do  not fit
into this pattern, and we may not be able to definitively 
classify these stars as either bulge or disk.  There is the
additional possibility that these stars originate from
a different stellar population altogether.

All calculations were carried out with the same methods employed
in the analysis of the metal-rich disk star \mul\ by
Castro  {\it et al.} (1996), so that our results can be scaled to this
prototype and extensively studied star.

In Sect. 2 the observations are described. In Sect. 3 the
detailed analysis is reported. In Sect. 4 the discussion and conclusions
are presented.

\section{Observations}

High-resolution spectra were obtained at the 2.1 m
Struve telescope of the McDonald Observatory, using 
the Sandiford \'echelle spectrograph at the Cassegrain focus
(McCarthy {\it et al.} 1993).
A two-pixel resolution of R = 60000 was obtained  with a
1.2 arcsec slit width, covering
the wavelength region $\lambda\lambda$ 6000-7780 ${\rm \AA}$,
contained in 30 orders. 
The slit length was 6.0 arcsec to prevent overlapping
orders at the long wavelength limit. The reductions were carried out by using IRAF. 
The \'echelle data were extracted
using an optimal background extraction routine (Tomaney \& McCarthy 1996).
The log-book of observations is
reported in Table 1 and a typical spectrum is shown in Fig. 1. 
The signal-to-noise (S/N) values reported in Table 1 are each a mean of several spectral
orders.

\section{Detailed Analysis}

The spectrum synthesis code used assumes Local Thermodynamic
Equilibrium (LTE) and is described in Cayrel {\it et al.} (1991).
The line list includes molecules of CN ${\rm A^2\Pi-X^2\Sigma}$,
${\rm C_2}$  ${\rm A^3\Pi-X^3\Pi}$ and TiO  ${\rm A^3\Phi-X^3\Delta}$
(see Barbuy {\it et al.} 1991; Milone {\it et al.} 1991).
Curves-of-growth were computed using the code RENOIR
by M. Spite.

For all calculations, the model atmospheres by Kurucz (1992)
were employed, where interpolations were made for the stellar
parameters of our sample stars.

The equivalent widths of the lines were measured using the
IRAF package and are reported in Tables 2a and 2b, together with
excitation potential $\chi_{ex}$ and oscillator strengths
log {\it gf}. The log {\it gf} values of \ion{Fe}{1} lines were adopted from 
Nave {\it et al.} (1994), while log {\it gf} values for elements other than
Fe were taken from several sources and most of them were then
fitted to the solar spectrum, as described in Castro {\it et al.} (1995).
Sources for log {\it gf} values of elements other than Fe are listed
in Table 2b. The selected \ion{Fe}{1} lines were chosen by discarding
those showing known blends. In principle, most of the \ion{Fe}{1}
lines are clean of blends.

The continuum adopted for the measurement of equivalent widths is very
close to the observed continuum and we did not attempt to place a
higher continuum to take into account blanketing effects. An example
of the continuum placement is given in Fig. 1.

\subsection{Temperatures}

The effective temperatures derived based on the Geneva photometry,
given in Table 3, were checked through excitation equilibrium
of \ion{Fe}{1} lines. A mean difference of 150 K above the values
derived from the Geneva
photometry were found, as also noticed by Cayrel de Strobel (1996). 
In Fig. 2a, the curve-of-growth of \ion{Fe}{1} for HD 138776 shows the 
good agreement between the lines of different excitation potential. 
Such temperatures were further
checked using the H$\alpha$ line profile which
confirmed the excitation equilibrium value. The H$\alpha$ profiles
were computed using a revised version of the code HYDRO by Praderie (1967).
In Fig. 2b is shown
the fit of the computed H$\alpha$ wings to the observed profile 
for HD 138776.
The final adopted temperatures are reported in Table 3.

\subsection{Gravities}

The gravities were derived through ionization equilibrium
of \ion{Fe}{1} and \ion{Fe}{2} lines. The values so derived were found to be
within 0.18 dex of the gravities predicted by the Geneva
photometry. An average of 10 \ion{Fe}{2} lines were available
(Table 2b), which makes the gravity determination
reliable.

\subsection{Metallicities}

The metallicities were calculated using \ion{Fe}{1} lines with W $<$ 150 m${\rm\AA}$,
in order to avoid saturation effects. Using all those \ion{Fe}{1} lines as
given in Table 2a, the metallicity derivation was carried out by
comparing the theoretical curve of growth to that for the measured
equivalent widths (Fig. 2a). The final metallicities are given in Table 3.

We also tested
 contamination by CN lines adopting a list selected by MR94 of CN-free
 \ion{Fe}{1} lines, but changes in metallicities for the sample stars
 are negligible; we emphasize that the effect of CN contamination is more
 important in giant stars of low temperatures and high metallicities. 
However, our data are 3 times the resolution of MR94 and
are generally higher S/N as well, making Fe lines in our data
much less likely to have been blended with CN.
 
For most stars the resulting [Fe/H] are in close agreement with the
values predicted by the Geneva photometry. The results
for BD-10$^o$3166, HD 99109, HD 115589, HD 126614 and HD 138776
are among the most metal-rich values known. For example, in
an analysis of 50 metal-rich stars of the Galactic disk, 
F95 obtained \fe $\approx$ 0.40 for the most
metal-rich ones of her sample.

Another interesting result is a comparison of our results to those
for \mul, considered the most metal-rich disk star,
and \bww, one of the strongest-lined stars in the Rich (1988) sample:
for these stars Castro {\it et al.} (1996) obtained \fe =
$+$0.46 and $+$0.47 respectively, using the same analysis
procedures employed in the present work.

\subsection{Microturbulent Velocities}

The microturbulent velocities were derived from the curves of growth
which best fitted the weak lines and the flat portion of the curves
at the same time. The microturbulent velocities derived for the
sample stars are listed in Table 3.

\subsection{Relative Abundances}

Table 4 gives the abundance ratios for analyzed elements relative to iron,
as discussed below. Figures 3a through 3d show the results for O, Na, Mg, Si;
Figures 4a through 4d give the results for 
Ca, Ti, V, Ni and Figures 5a through 5d show the abundance pattern for Y, Zr, Ba and Eu in the sample stars. The linear fits of
[element/Fe] vs. [Fe/H] are indicated in the figures.

{\it Sodium-to-Iron}: [Na/Fe] ratio is enhanced and increases with
increasing metallicity in agreement with F95. This would suggest an
additional source of Na enrichment in the Galaxy such as synthesis of
Na in the hydrogen burning shell of intermediate mass stars (Timmes,
Woosley \& Weaver 1995).

{\it Oxygen}: we have fitted synthetic spectra of the forbidden line [OI] at ${\rm \lambda}$6300.311 ${\rm \AA}$ 
by taking into account CO association. Since we do not dispose of carbon
lines, we assume [C/Fe]=$-$0.2, 0.0 and $+$0.2; the resulting oxygen
abundance is not much affected by the carbon abundance in these hot
stars, as indicated in Table 5.
We found that oxygen
follows the trend of [O/Fe] ratio in the disk: oxygen abundances in our sample of stars declines from the solar value
towards the most metal-rich stars and reaches [O/Fe]=$-$0.23, following
the behavior of data by Nissen \& Edvardsson (1992). Our stars seem to be
the upper limit of [Fe/H] in their scale.

{\it $\alpha$-elements Mg, Si}:
Spectrum synthesis of the \ion{Mg}{1} line at ${\rm
\lambda 6319.242}$ ${\rm \AA}$ was employed to derive Mg
abundances.  The
\ion{Si}{1} line at ${\rm \lambda 6721.84}$ used to derive silicon
abundances is a reliable abundance indicator as claimed by Gratton
\& Sneden (1990) 
(note that MR94 found this feature to be blended and
to give spuriously high abundances in metal-rich stars, less
of a concern at our resolution). Mg and Si relative to Fe are approximately constant
with metallicity.
The average of Mg abundance for our sample is about
${\rm \approx -}$0.06 and
[Si/Fe] slightly increases with increasing metallicity.

{\it $\alpha$-elements Ca, Ti}: calcium and titanium are essentially solar to subsolar
relative to Fe. Abundances derived from \ion{Ca}{1} lines at ${\rm \lambda 6166.440}$, ${\rm \lambda 6455.605}$, ${\rm \lambda 6508.846}$ ${\rm \AA}$ and \ion{Ti}{1} line at ${\rm \lambda 6336.113}$ ${\rm \AA}$ show a slight decline
as metallicity increases, the same result found by Edvardsson {\it et al.}
(1993).  The \ion{Ti}{1} line used for spectrum synthesis is 
shown in Fig. 6 for HD 115589. It is a weak and the most reliable line
of our list (the same was observed by MR94).

{\it Iron-peak-elements V, Ni}: vanadium and nickel behave in different ways. 
V is overabundant with a mean value of $+$0.2 dex,
%
but may be affected by hyperfine splitting. Ni shows a slight increase
towards the more metal-rich stars and both results are in good
agreement with the work by Porto de Mello (1996) who analyzed the
abundance distribution of solar-type stars in the solar neighborhood.

{\it r-process-element Eu}: europium was available for 6 sample stars. 
The solar
ratio is found for Eu in 5 stars, and G 161-29 
shows [Eu/Fe]=$+$0.5. This result for G 161-29 would imply that Supernovae Type II
would have had importance in the chemical enrichment of its original
gas (Meyer 1994). This star will be discussed at the end of this section.

{\it s-process-elements Y, Zr, Ba}: We have found that Y, Zr and
Ba are underabundant relative to iron. Y and Ba tend to decrease to
subsolar values as metallicity increases, again in agreement with 
F95 and Porto de Mello (1996). On the other hand, Zr is subsolar
in all sample stars in disagreement with most stars of F95's sample
but compatible with her values for the coolest stars. The two entries
for the \ion{Ba}{2} line in Table 2b are hyperfine components of a
same line as described by Fran\c cois (1996).

The s-process abundances are clearly different from the sample of
MR94 and we discuss the issue further in Sec. 4.  Further study 
of the s-process elements in both these disk stars and Galactic
bulge giants is very important.

G161-29 seems to be different from the other sample stars. It shows high europium, titanium and calcium, but low magnesium and oxygen abundances
relative to iron: [Eu/Fe]=$+$0.5, [Ti/Fe]=$+$0.35, [Ca/Fe]=$+$0.2, 
[Mg/Fe]=$+$0.05 and [O/Fe]=$+$0.0.
In fact, this star might belong to the bulge population, showing similarities to
the MR94 results, where some  $\alpha$-elements
showed to be overabundant but not all of them. In Figure 7 is shown
([Mg/Fe]+[Ti/Fe])/2 for the MR94 bulge stars and our sample, where it
can be seen that G161-29 fits the bulge pattern.
It is also interesting to notice that G161-29 might be very similar to the
stars in the sample of Barbuy \& Grenon (1990) which show kinematic and
photometric properties of possible bulge members.
 
\subsection{Uncertainties}

We explore the effect of the following changes in input parameters:
${\rm \Delta \teff =\pm 100}$ K, ${\rm \Delta \logg = \pm 0.1}$ ${\rm cm.s^{-2}}$, 
${\rm \Delta \xi = \pm 0.4}$ ${\rm km.s^{-1}}$ and ${\rm \Delta W =
\pm 15}$
${\rm m\AA}$. The
dependence of metallicity on \teff\ and \logg\ is nearly negligible
and is $\pm$0.05. On the other hand, the variation of metallicity with
microturbulent velocity is ${\rm \Delta \fe}$=$-$0.10 dex as we apply the
change ${\rm \Delta \xi}$=+0.40 ${\rm km.s^{-1}}$.

The abundance ratio uncertainty has a random component which arises
from equivalent width measurement and is typically ${\rm \Delta \fe
\approx 0.10}$ dex. For an element X represented by several lines, the
random component of the uncertainty is diminished by the square root
of the number of lines. The typical mean uncertainty of our equivalent
width measurement is about 10 m${\rm \AA}$. 
We estimate a final uncertainty of about $\pm$0.2 in elemental
ratios [X/H].

\section{Conclusions}
	
We have acquired high-resolution, high S/N spectra of 9 G-K dwarf stars in the solar neighborhood, kinematically and photometrically selected by M. Grenon.
We have used curve of growth and spectrum synthesis calculations to
derive abundances for these stars.  We conclude that 5 of these
stars are now the most metal-rich stars with a high-resolution
abundance analysis.

The abundance pattern of our sample 
stars, illustrated in Fig. 8 where the elemental ratios
[X/Fe] are plotted vs. the atomic number Z, is characterized by:

(i) Sodium-to-iron ratio is overabundant for all stars where this element
was available.

(ii) There is a decrease of the $\alpha$-elements-to-iron (O, Ca, Ti)
with increasing metallicity, as illustrated in Fig. 9a which compares
the mean  $\alpha$-element abundance given in Table 12 of Edvardsson
{\it et al.} (1993), and the mean value of our determination for Mg, Si, Ca and Ti. Another
interesting comparison is given in Fig. 9b which shows the [O/Fe] data of 
Nissen \& Edvardsson (1992)'s stars plotted with our data.
 
(iii) Europium-to-iron ratio is about solar for all stars except for G161-29
(notice that this star also shows enhancement of Ti and Ca). It is
interesting to compare the [Eu/Fe] ratio for our 5 disk stars to the
results obtained by Woolf, Tomkin \& Lambert (1995). The trend of
decreasing [Eu/Fe] with increasing metallicity shown by their data
is not confirmed by the [Eu/Fe]${\rm \approx}$ 0.0 of our sample stars
as given in Fig. 9c.

(iv) s-process elements are underabundant relative to iron.

In comparison to MR94, the most striking difference is that
the s-process elements are subsolar here, which was not the
case in MR94.  We note that [Ba/Fe] in particular can
be used as a crude chronometer (Edvardsson {\it et al.} 1993) such
that the lower [Ba/Fe], the older the stars.  A straightforward
application leads us to conclude that these stars must be
older than those of MR94.  On the other hand, the clearly
established tendency for [O/Fe] to continue declining at
high metallicities argues for a substantial contamination
by Type I SNe; i.e., the population formed over so long
a timescale that the slower Type I SNe could be the dominant
source of metals.

In principle, the decrease of O, Ca, Ti, Y and Ba 
relative to Fe with increasing
metallicity could also be explained by infall of material processed in
halo Supernovae Type I, as suggested by Freitas Pacheco (1993).
However this is contradicted by the high [Mg/Fe] $\approx$ 0 and
[Si/Fe] $\approx$ 0 in the metallicity
range of our sample, which shows that 
a more complex scenario is necessary to explain the chemical
enrichment of the most metal-rich disk stars.

\subsection{The Nature of these Stars}

Are our stars members of the local disk, interlopers from the
bulge, or an entirely new stellar population?

The location of our sample stars in the color-magnitude diagram 
${\rm M_v}$ vs. ${\rm (B_2-V_1)}$ of the most metal-rich stars in the solar neighborhood within 120 pc is shown
in Fig. 10. The absolute magnitudes of our sample stars derived from
the Geneva photometry and those derived from the Hipparcos parallaxes
are indicated by circles and filled squares respectively. Our
sample stars define a locus of an evolved  main-sequence composed
of the most metal-rich stars.   We first consider the possibility
that these stars are the final stages of the evolution of the disk.
If this is the case, their high eccentricity orbits and chemistry
support their ancient nature, and one might speculate that
these stars represent the final generation of the inner disk population.

Rose (1985) identified a class of strong-lined dwarf stars in the
Galactic disk using a gravity-independent metal-abundance index.
This index indicated that the stars were strong-lined in their Fe
lines and in general they appeared to be as strong-lined as the most
metal-rich stars in the Cayrel de Strobel \& Bentolila (1983) [Fe/H]
catalog, i.e., those stars with published [Fe/H]$>+$0.25.

Boulade, Rose \& Vigroux (1988) found that the three strong-lined G
dwarfs of their sample (which includes HD 126614 with [Fe/H] given
by Grenon), had near UV indices that clearly established them to be
more strong-lined than solar-abundance stars. They hence concluded
that although there was no detailed abundance analysis carried
out for those unusual G dwarfs at that time, their strong-lined
nature was evident in a variety of spectral indices.

Membership in the Galactic bar is attractive, but these data do not
provide a confirmation.  Since a bar dominates the central kpc,
it is possible that some stars on chaotic orbits
manage to diffuse out of, or are ejected from, the bar population.
Thus the local Grenon stars would be these members of the bar
that were ejected.

The similarity to the MR94 sample is tantalizing (Figure 7) but
not complete.  As their metallicity increases, the composition
of the MR94 bulge giants tends toward Solar; this is also
true of the s-process elements.  While the $\alpha-$element
patterns look very similar to the metal-rich bulge giants
of MR94, the clearly subsolar
s-process abundances are a problem and make
the bulge origin look much less likely.
Because they are produced in the envelopes of AGB stars, s-process elements 
are likely enhanced on a longer timescale than that 
of the Fe-peak elements produced
in Type I SNe (and more easily distributed in the ISM).

The final possibility is that the stars originate in a new stellar
population, perhaps the older, central regions of the disk.  
While the deep potential well of the Galactic center would
appear to be the logical place for stars to reach high metallicities,
it appears that this process has occurred in surprising places
such as old Galactic clusters.  It is conceivable that the early
disk was able to reach very high local abundances even at large
galactocentric radii.  We are led to conflicting interpretations:
ancient origins from the s-process, color-magnitude diagram, and
enhanced alpha-elements, yet an extended formation history and
Type I SN enrichment based on the decline of [O/Fe] to subsolar
values in the most metal-rich stars.

While the nature of these peculiar stars remains uncertain,
there is a clear need for a larger sample of these high-velocity,
metal-rich stars.

\acknowledgements

We are grateful to Greg Langmead for the
data reduction. SC acknowledges the Fapesp PhD fellowship
n$^o$ 92/1351-3 and the CNPq fellowship n$^o$ 201703/93-9.
We acknowledge useful discussions with Andy McWilliam,
James Truran, and George Wallerstein.

\newpage
\centerline{FIGURE CAPTIONS}
\parskip 2mm
\parindent 0mm

\figcaption{\label{figi} HD 99109:
a typical spectrum showing the high S/N ratio and the local continuum.}

\figcaption{\label{figii}
(a) Curve-of-growth of \ion{Fe}{1} for HD 138776, illustrating the agreement
between lines of different excitation potential. (b) H$\alpha$ line in 
HD 138776: observed spectrum (solid line)
and synthetic spectra computed for \teff = 5500 K (dotted line)
and 5700 K (dashed line) which represents the best fit.
} 

\figcaption{\label{figiii}
Abundance ratios vs. \fe\ for our sample stars for: (a) [O/Fe];
(b) [Na/Fe]; (c) [Mg/Fe]; (d) [Si/Fe].
 Linear fits
[element/Fe] vs. [Fe/H] are indicated in the diagrams.
}

\figcaption{\label{figiv}
Abundance ratios vs. \fe\ for our sample stars for: (a) [Ca/Fe]; (b) [Ti/Fe]; (c) [V/Fe]; (d) [Ni/Fe].  Linear fits
[element/Fe] vs. [Fe/H] are indicated in the diagrams.
}

\figcaption{\label{figv}
Abundance ratios vs. \fe\ for our sample stars for: (a) [Y/Fe]; (b) [Zr/Fe]; (c) [Ba/Fe]; (d) [Eu/Fe].  Linear fits
[element/Fe] vs. [Fe/H] are indicated in the diagrams.
}

\figcaption{\label{figvi}
Synthetic spectrum of the \ion{Ti}{1} line at ${\rm \lambda 6336.113}$
 ${\rm \AA}$ for HD 115589. The lines represent the synthetic spectra calculated
with [Ti/Fe] = $-$0.3 (dotted), 0.0 (solid), +0.3 (dashed) and the observed
spectrum (crosses).
}

\figcaption{\label{figvii}  
 [${\rm \alpha}$/Fe] vs. [Fe/H] for McWilliam \& Rich (1994) bulge
stars (triangles) and the sample stars (squares) where only Mg and
Ti are taken into account. ${\rm <Mg+Ti>={1\over 2} ([{Mg\over Fe}]+
[{Ti\over Fe}])}$
}

\figcaption{\label{figviii}
Element ratios relative to iron [X/Fe] for the sample stars
vs. the atomic number Z: upper points (shifted by a constant 
equal to 1.0 in [X/Fe]) correspond to the 5 stars
with [Fe/H] $\geq$ 0.40 (open circles) and lower points 
refer to the 4 stars with [Fe/H] $\leq$ 0.30 (filled circles).
}

\figcaption{\label{figix} 
(a) [${\rm \alpha}$/Fe] vs. \fe\ for the sample stars 
(squares) compared to
those from Edvardsson {\it et al.} (1993) (x),
where [${\rm \alpha}$/Fe]= ${\rm {1\over 4} ([{Mg\over Fe}]+
[{Si\over Fe}]+[{Ca\over Fe}]+[{Ti\over Fe}])}$;
(b) [O/Fe] vs. \fe\ for our sample stars (squares) compared to
those from Nissen \& Edvardsson (1992) (triangles); (c) [Eu/Fe] vs. \fe\ for our sample stars (squares) compared to
those from Woolf, Tomkin \& Lambert (1995) (x).
}

\figcaption{\label{figx}
Color-magnitude diagram of our sample stars plotted
together with solar neighborhood 
main sequence stars. The absolute magnitudes
are derived from the Geneva photometry (circles) and from the
Hipparcos parallaxes (filled squares).
}

\end{document}